\documentstyle[epsfig,float,multicol,prb,aps]{revtex}

\begin{document}
\ifx\href\undefined
\else
\errmessage{Don't use hypertex}
\fi

\title{Nonlocal density functionals and the linear response of
 the homogeneous electron gas }
\author{I.I. Mazin$^{1}$ and D.J. Singh$^{2}$}
\address{$^1$Code 6691, Naval Research Laboratory, Washington, DC 20375\\
$^2$CSI, George Mason University, Fairfax, VA}
\date{\today}
\maketitle

\begin{abstract}
The known and  usable truly nonlocal functionals for
exchange-correlation energy of  the inhomogeneous electron gas are the
 ADA (average
density approximation) and the  WDA (weighted density approximation). ADA, by
design, yields the correct linear response function of the uniform electron gas.
WDA is constructed so that it is exact in the limit of 
one-electron systems. We derive an expression for the linear response of the
uniform gas in  the WDA, and calculate it for several flavors of WDA. We
then compare the results with the Monte-Carlo data on the
exchange-correlation local field correction, and identify the weak points of
conventional WDA in the homogeneous limit. We suggest how the
WDA can be modified
to improve the response function. The resulting
approximation is a good one in both opposite limits, and should be 
useful for practical nonlocal density functional calculations.
\end{abstract}

\begin{multicols}{2}
     Calculations based on the Kohn-Sham formulation of density functional
theory\cite{IEG}
have become a prominent tool in condensed matter physics. Current work
is dominated by local density approximation (LDA) studies, in which
the exchange correlation functional is a local function of the density.
However, as the number and accuracy of calculations has increased,
so has
the number of well documented cases where the LDA is inadequate
and with this interest in beyond LDA approaches, {\it e.g.,}
the generalized gradient approximation (GGA), which depends
locally on both the density and its gradient.

     Modern GGA functionals do 
improve upon LDA results for a wide range of problems.
However, several studies have pointed out deficiencies in GGA
functionals, e.g. difficulties in describing ferroelectric materials, and
cases of overcorrection of LDA errors particularly in materials containing 
heavy atoms.  Both  the LDA and GGA fail to
provide a correct description of  the 
static short-range linear response of the
homogeneous electron gas.
All this leads to the question of the extent to which truly non-local
functionals are practical and able to correct the deficiencies of LDA and
GGA methods.

The first efforts at developing practical non-local functionals date
from the 1970's when the weighted density approximation (WDA) and average
density approximation (ADA) were proposed. However, over most of the
intervening period the field has been relatively dormant, in part because
of the success of the simpler LDA and GGA schemes and in part because it
was widely thought that such schemes could not be implemented 
in a computationally tractable fashion. However, at least for the WDA,
computationally efficient algorithms are now known \cite{Singh,tet,char}
and benchmark calculations have been reported.
In the cases that
have been studied ground state properties are generally improved over the
LDA\cite{char1}.

Both methods were proposed in 1978--1979 by Alonso {\it et al} and
Gunnarsson {\it et al}\cite{AG,GJL}. Both exploit the general expression for 
$E_{xc},$ 
\begin{equation}
E_{xc}=\frac{e^{2}}{2}\int \frac{n({\bf r})n({\bf r}^{\prime })}{|{\bf r-r}%
^{\prime }|}G({\bf r,r}^{\prime })\{n({\bf r)}\}d{\bf r}d{\bf r}^{\prime },
\label{exc}
\end{equation}
where the function $G({\bf r,r}^{\prime })$ is also a functional of the
total electronic density $n{\bf (r).}$ A rigorous expression for $G$ can be
derived\cite{IEG} in terms of coupling constant averaged
 pair correlation function:
 $G({\bf r,r}^{\prime })=\int_{0}^{1}[g(%
{\bf r,r}^{\prime };\lambda )\{n({\bf r)}\}-1]d\lambda .$ For the uniform
gas this function, $G_{0}(|{\bf r-r}^{\prime }|,n),$ is known with high
accuracy\cite{PW}, but for an arbitrary system there is no practical way to
use this formula. 
The LDA instead of Eq. (\ref{exc}) uses $(e^{2}/2)\int d{\bf r}%
d{\bf r}^{\prime }n^{2}({\bf r})G_{0}[|{\bf r-r}^{\prime }|,n({\bf r)]}/|%
{\bf r-r}^{\prime }|,$ so that $E_{xc}$ becomes $E_{xc}^{LDA}=\int n({\bf r}%
)\epsilon _{xc}[n({\bf r)]}d{\bf r,}\ \ \epsilon _{xc}$ being the density of
exchange-correlation energy of the uniform gas.  The
LDA is incorrect in the two
important limits:  the fully localized, {\it i.e.}, a one electron
system, and  the fully delocalized limit, {\it i.e.}, homogeneous electron gas.
In the former case
the LDA
gives a spurious self-interaction with  energy $(e^{2}/2)\int d%
{\bf r}d{\bf r}^{\prime }n({\bf r})n({\bf r}^{\prime })/|{\bf r-r}^{\prime
}|+\int n({\bf r})\epsilon _{xc}[n({\bf r)]}d{\bf r},$ which is widely thought 
\cite{sic} to be a key problem with  the LDA. In the
homogeneous limit, the  LDA gives the  correct exchange-correlation
energy, but the {\it changes} of this energy upon small perturbations
are not properly described;  the  second variation of $E_{xc}$ with
density, {\it i.e.}, the  exchange-correlation part of the dielectric response, 
$K_{xc}({\bf r-r}^{\prime })=\delta ^{2}E_{xc}/\delta n({\bf r})\delta n(%
{\bf r}^{\prime }),$ is a delta function, which is incorrect. The Fourier
transform of $K_{xc}(r)$ in LDA is independent of the wave vector. 
Since LDA is exact for the uniform gas, $K_{xc}^{LDA}$
corresponds to the correct $K_{xc}$ at $q=0.$
GGAs also give correct behavior at $q=0$, but become even worse than  the
LDA at high  $q$'s.

The two nonlocal expressions for $E_{xc}$,
 WDA and ADA, aimed at correcting one or the other of these two
limits.
The former uses the general expression (\ref{exc}), but instead of
the actual function $G$ uses a model function, defined so
that the one electron limit is honored. 
This begins by choosing a generic expression for $G,$ which depends on one
parameter $\bar{n},$ to be defined later. In the original papers it was
suggested that $G({\bf r,r}^{\prime },\bar{n})=G_{h}({\bf r,r}^{\prime },%
\bar{n})=\int_{0}^{1}[g({\bf r,r}^{\prime };\lambda ,\bar{n})-1]d\lambda .,$
where $g$ is the pair correlation function of the homogeneous electron gas.
Later it was realized\cite{GJ} that other choices of $G$ may be better than $%
G_{h}.$ In the WDA $\bar{n}$ is a function of ${\bf r,}$ but differs
from $n({\bf r),}$ and is chosen so that $\int G[{\bf r,r}^{\prime },\bar{n}(%
{\bf r})]n({\bf r}^{\prime }{\bf )}d{\bf r}^{\prime }=-1.$ This assures that
for a one electron system $E_{xc}$ cancels the self-interaction
exactly.

In the  ADA  $n({\bf r}^{\prime })$ in Eq. (%
\ref{exc}) is substituted by $n({\bf r}),$ which results in $%
E_{xc}^{ADA}=\int n({\bf r})\epsilon _{xc}[\tilde{n}({\bf r)]}d{\bf r.}$
Then $\tilde{n}({\bf r)}$ is defined as $\tilde{n}({\bf r)=}\int w[|{\bf r-r}%
^{\prime }|,\tilde{n}({\bf r})]n({\bf r}^{\prime }{\bf )}d{\bf r}^{\prime },$
and the universal function $w$ is chosen so that $\delta
^{2}E_{xc}^{ADA}/\delta n({\bf r})\delta n({\bf r}^{\prime })$ gives the
correct $K_{xc}$ for the uniform gas. Contrary to the WDA, the
ADA  is not self-interaction free in one electron systems.

From the beginning there was substantial
interest in the  behavior of WDA in the delocalized limit
\cite{IEG}. Williams and von Barth \cite{IEG(WB)}
suggested that the WDA should give substantial improvement over the  LDA in this
limit, but till now no systematic study has been reported. If this
conjecture is true, the WDA has a great advantage over any other known
approximation to the DFT in the sense that it accurately reproduces two key
physical limits. Furthermore, even if it is not entirely correct,
the next question is, whether or not an approximation based on the WDA exists
that does provide proper limiting behavior. In this paper we derive an
expression for $K_{xc}$ in the WDA, calculate $K_{xc}$ for popular flavors of
WDA, and discuss construction of an improved WDA method.

We start by deriving a closed expression for $K_{xc}$ in the WDA for
an arbitrary $G.$ First some 
notation: we denote the product $(e^{2}/r)G(r)$ as $W(r),$ use atomic units
where $e=1,$ $\hbar =1,$ and use primes for the derivative with respect to
the density argument, {\it e.g.} $G^{\prime }=dG/dn.$ We also
introduce two functions, reflecting implicit dependence of the weighted
density $\bar{n}$ on variations of the real density: 
\begin{eqnarray}
d({\bf r}^{\prime }{\bf -r}) &=&\delta \bar{n}({\bf r}^{\prime })/\delta n(%
{\bf r}) \\
f({\bf r}^{\prime }{\bf -r,r}^{\prime }{\bf -r}^{\prime \prime }) &=&\frac{%
\delta ^{2}\bar{n}({\bf r}^{\prime })}{\delta n({\bf r})\delta n({\bf r}%
^{\prime \prime })}=\frac{\delta d({\bf r}^{\prime }{\bf -r})}{\delta n({\bf %
r}^{\prime \prime })}.
\end{eqnarray}
Using the WDA expression for the exchange-correlation energy, 
\begin{equation}
E_{xc}=(1/2)\int n({\bf r})n({\bf r}^{\prime })W[|{\bf r-r}^{\prime }|,\bar{n%
}({\bf r)]}d{\bf r}d{\bf r}^{\prime },
\end{equation}
we can express $K_{xc}$ in terms of functions $d$ and $f,$ and we can find
these functions using normalization condition 
\begin{equation}
\int d{\bf r}^{\prime }n({\bf r}^{\prime })g[|{\bf r-r}^{\prime }|,\bar{n}(%
{\bf r})]=-1,
\end{equation}
and the LDA limit condition 
\begin{equation}
\int d{\bf r}^{\prime }W[|{\bf r-r}^{\prime }|,n]=2\epsilon _{xc}/n.
\end{equation}
We proceed in reciprocal space, which corresponds to
using density perturbation of the form $\delta n({\bf r)=}n_{q}e^{i{\bf qr}}$%
. Let $W_{q},$ $G_{q},$ $d_{q}$ and $f_{p,q}$ will be the Fourier transforms
of the corresponding functions. Then the above condition can be written as 
\begin{equation}
G_{0}=-1/n,\qquad W_{0}=2\epsilon _{xc}/n.
\end{equation}
For $d_{q}$ one finds $d_{q}=-nG_{q}.$ For $f_{p,q}$ we need only diagonal
elements, $f_{q,-q}=2nG_{q}(nG_{q}^{\prime }+G_{q}).$ In terms of $d$ and $%
f, $ $K_{xc}$ is 
\[
K_{xc}(q)=W_{q}+n_{0}d_{q}W_{q}^{\prime }+n_{0}d_{q}W_{0}^{\prime
}+{n_{0}^{2}\over 2}
(d_{q}^{2}W_{0}^{\prime \prime }+n_{0}^{2}f_{q,-q}W_{0}^{\prime
}).
\]
resulting in
\begin{equation}
K_{xc}(q)=W_{q}-n^{2}G_{q}(W_{q}^{\prime }+W_{0}^{\prime
})+n^{2}(n^{2}G_{q}^{2}W_{0}^{\prime })^{\prime }/2.  \label{kxcWDA}
\end{equation}

\begin{figure}
\centerline{\epsfig{file=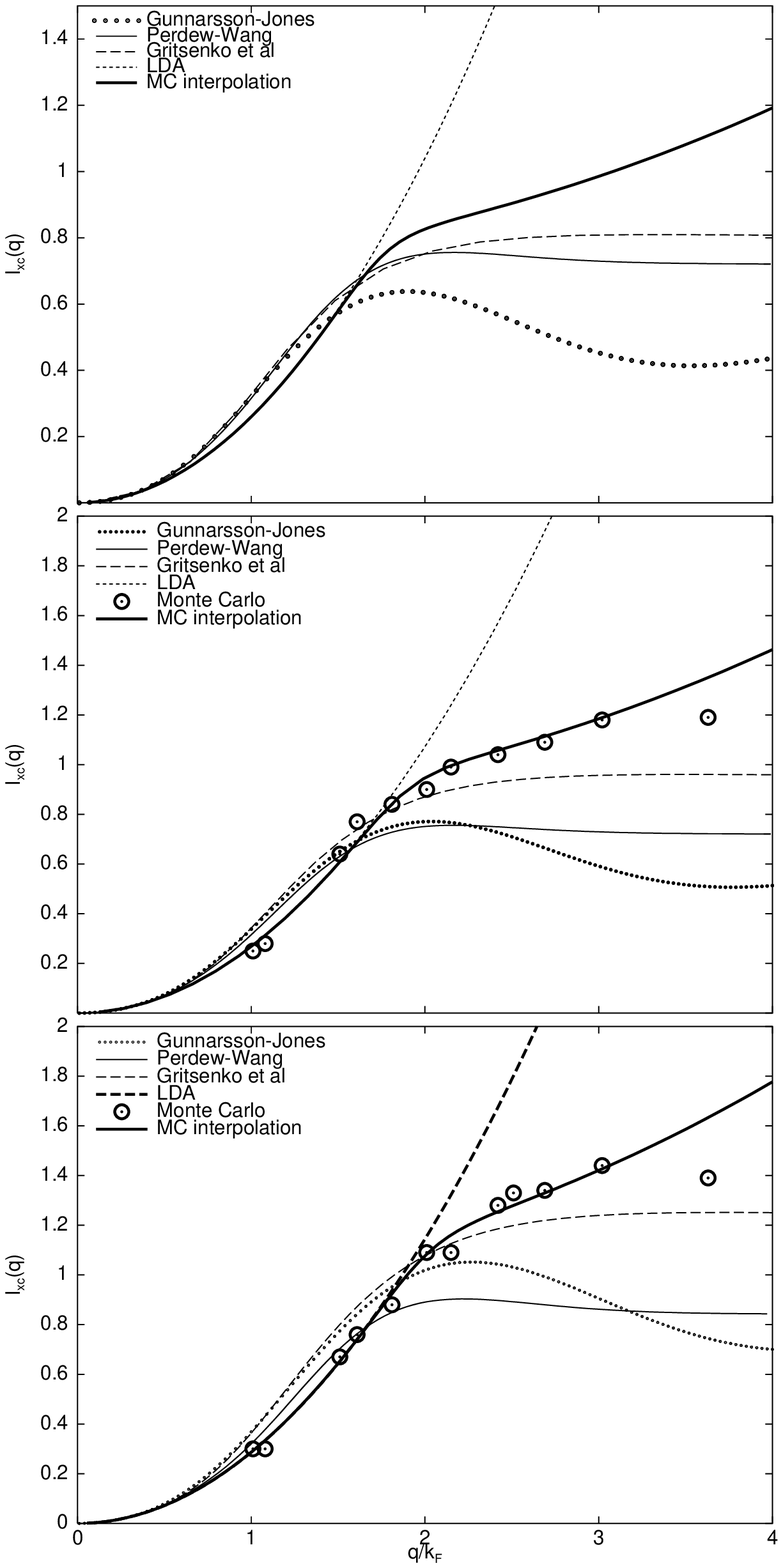,width=0.9\linewidth}}
\vspace{0.1in} \setlength{\columnwidth}{3.2in} \nopagebreak
\caption{Exchange-correlation local field factor in the WDA of
Ref.\protect\onlinecite{GJ} (Gunnarsson-Jones),
Ref.\protect\onlinecite{Gri} (Gritsenko et al), and derived from
the homogeneous electron gas pair correlation function (Perdew-Wang),
as compared with the Monte Carlo results (Monte Carlo)
and the interpolating formula thereof (MC interpolation), as 
given in  Ref.\protect\onlinecite{MC}. Densities, from top to bottom,
correspond to $r_s=1,2,5$.}
\end{figure}
The original formulation of the WDA used the corresponding homogeneous
electron gas function for $G$. Since then, three forms of $G$ have 
been used in the calculations, all of which result in improvement over LDA
(in the admittedly limited number of tests performed to date).
These are: the function $G$ derived for the uniform gas by Perdew and
Wang\cite{PW}, the Gunnarsson-Jones function $%
G^{GJ}(r)=C_{1}(n)\{1-\exp [-\left( \frac{r}{C_{2}(n)}\right) ^{-5}]\},$ and
 the Gritsenko
et al \cite{Gri} function $G^{GRBA}(r)=C_{1}(n)\exp [-\left( \frac{r%
}{C_{2}(n)}\right) ^{k}]\},$ $k=1.5$ (note that the uniform gas function\cite
{PW} is approximately given by the same expression with $k=2$). We tested
these functions for the densities $r_{s}=1,$ $2,$ and 5 and obtained modest
agreement with the Monte Carlo results\cite{MC} (Cf. Fig.1, where we
plotted calculated exchange-correlation local field factor $I_{xc}(q)=\frac{%
q^{2}}{4\pi }K_{xc}(q),$ and compare it with Monte Carlo data\cite{MC}). By
construction, $K_{xc}(0)$ is correct (and is in fact the LDA value). At $q%
\agt1.5-1.8k_{F}$ $K_{xc}$ falls below its LDA value and continues to
decrease at large $q$'s. However, a closer look reveals two major
disagreements: first,
$I_{xc}^{WDA}(q)$ is considerably larger than the
Monte-Carlo data for the wave vectors between $\approx 0.5k_{F}$ and $%
1.5k_{F}$. Second, $I_{xc}(q)$ in WDA tends to a
constant value equal to $\lim_{q\rightarrow \infty }[W_{q}/(4\pi /q^{2})].$
In Monte Carlo calculations it is $K_{xc}(q)$ itself that has a finite limit
at $q\rightarrow \infty ,$ while $I_{xc}(q)\rightarrow const\cdot q^{2}$ at $%
q\rightarrow \infty .$ The latter result was predicted by Holas \cite{Holas}
and is physically important: it reflects the fact that $E_{xc}$ is not
solely an interaction energy, but also includes the exchange-correlation
contribution to kinetic energy (which in fact decays slower with $q$ than
the interaction part of $E_{xc}).$

Can one correct these two deficiencies without
compromising the correct one-electron limit of WDA? In fact, it
was noticed long ago\cite{GJ} that there is no particular reason to use
the homogeneous electron gas pair correlation function for $G.$ Since using $%
G_{h}$ in WDA does not guarantee any improvement in describing
properties of the homogeneous gas itself, one may use the
freedom in $G(r)$ to adjust the WDA so that the calculated
local field factor (and thus linear response function) is as accurate as
possible. Inversion of
eq.(\ref{kxcWDA}) yields $G(q)$
for a given $K_{xc}(q).$ It does not guarantee, however, that
the result will be physical. So, as a first step, let us analyze Eq. (\ref
{kxcWDA}). For this purpose, we  write $%
G_{q}=-\varphi (p/Q)/n$, with the condition $\varphi (0)=1,$ where $Q$ is
some constant (both the Gunnarsson-Jones and the 
Gritsenko {\it et al} functions are of
this form). Then 
\begin{eqnarray}
W_{p} &=&\frac{1}{8\pi ^{3}}\int d^{3}q\frac{4\pi }{|{\bf q-p|}^{2}}G_{q}=%
\frac{1}{\pi p}\int_{0}^{\infty }qdq\log \left| \frac{q+p}{q-p}\right| G_{q}
\label{Wp1} \nonumber \\
W_{0} &=&\frac{2}{\pi }\int_{0}^{\infty }G_{q}dq=-\frac{2Q}{\pi n_{0}}%
\int_{0}^{\infty }\varphi (x)dx=\frac{2\epsilon _{xc}(n_{0})}{n_{0}}. 
\nonumber
\end{eqnarray}

If we now define $Q(n)=-\pi \epsilon _{xc}(n),$ then the second condition on 
$\varphi (x)$ becomes $\int_{0}^{\infty }\varphi (x)dx=1.$ These two
conditions reduce our freedom to adjust $G_{q}:$ since the
characteristic size of $\varphi (x)$ is of order of 1, the wave vector
dependence of $G_{q}$ is defined by the ratio $q/Q=-q/\pi \epsilon _{xc}.$
At high density $Q=1.333k_{F},$ and only close to $r_{s}\agt6$ does
it approach
$2k_{F},$ the number at which real local field factor changes its behavior
from low-$q$ to the high-$q$ limit. A monotonic function $\varphi (x)$ does
not reproduce this feature, which explains why existing WDA parametrizations
put the hump in $K_{xc}$ at  too low $q.$ Nonmonotonic and explicitly
density-dependent functions $\varphi (x)$ may be able to shift the hump
to its correct position at $q=2k_{F}.$ It is still an open
question whether or not a physically sound function can be found with 
this property. 

However, even if the ``$2k_{F}$'' problem is fixed, another, probably
even more important problem remains: the short wave length behavior of $%
K_{xc}.$ Fortunately, this is easy to correct. Farid {\it et al}\cite{farid}
tabulated the coefficient $\gamma $ that defines the asymptotic behavior of $%
K_{xc}(q\rightarrow \infty )$ as $K_{xc}(q\rightarrow \infty )=-\frac{4\pi }{%
q^{2}}\gamma (n)\frac{q^{2}}{k_{F}^{2}}.$ These values can be fit
as 
$$
\gamma (n) =(\frac{9\pi }{4})^{4/3}\frac{f(\sqrt{r_{s}})}{15}\ \ \ \ 
f(x) ={x(a+bx)\over (1+cx+dx^{2})}
$$
where
$a=0.026319,$ $b=0.00823859,$ $c=-0.173199,$ $d=0.233081$. 
Let us now modify the function $G(r)$%
\[
G(r)=G_{1}(r)+G_{2}(r)=A\delta (r)/4\pi r+G_{2}(r), 
\]
Since $\int G_{1}(r)r^{2}dr=0,$ the normalization condition for $G_{2}$ is
the same as for $G$ itself. Since $4\pi \int G_{1}(r)rdr=A,$ the LDA
limit condition for $G_{2}$ becomes 
\begin{eqnarray*}
4\pi \int G_{2}(r)rdr &=&2\tilde{\epsilon}_{xc}(n)/n, \\
\tilde{\epsilon}_{xc}(n) &=&\epsilon _{xc}(n)-An/2.
\end{eqnarray*}
Thus
\begin{eqnarray*}
A =-{4\pi \gamma (n)\over k_{F}^{2}}
&=&-(\frac{9\pi }{4})^{2/3}\frac{4\pi r_{s}^{2}%
}{15}f(\sqrt{r_{s}})=-3.0856r_{s}^{2}f(\sqrt{r_{s}}), \\
\tilde{\epsilon}_{xc}(n) &=&\epsilon _{xc}(n)+\frac{0.368317}{r_{s}}f(\sqrt{%
r_{s}})
\end{eqnarray*}
Now $G_{p}=G_{2p},$ $W_{p}=A+W_{2p},$ and $W_{p}^{\prime }=A^{\prime
}+W_{2p}^{\prime },$
\end{multicols}
\vskip -15pt \rule[1pt]{0.45\columnwidth}{.1pt} 
\begin{equation}
K_{xc}(q) =A+W_{2q}-n^{2}G_{2q}(A^{\prime }+W_{2q}^{\prime
}+W_{2,0}^{\prime })+n^{2}(n^{2}G_{q}^{2}W_{2,0}^{\prime })^{\prime }
\label{Kxcd} 
=A-n_{0}^{2}G_{2p}A^{\prime }+\tilde{K}_{xc},  \nonumber
\end{equation}
where $\tilde{K}_{xc}$ is calculated from $\tilde{\epsilon}_{xc}$ in exactly
the same way as $K_{xc}$ is calculated from $\epsilon _{xc}.$ The
corresponding functional for the exchange-correlation energy  is
\begin{equation}
E_{xc}^{AWDA}=\frac{1}{2}\int \frac{n({\bf r})n({\bf r}^{\prime })}{W(|{\bf %
r-r}^{\prime }|)}G[|{\bf r-r}^{\prime }|,\bar{n}({\bf r)]}d{\bf r}d{\bf r}%
^{\prime }+\int n({\bf r})\frac{0.368317}{\bar{r}_{s}({\bf r)}}f\left[ \sqrt{%
\bar{r}_{s}({\bf r)}}\right] d{\bf r}.  \label{awda}
\end{equation}
\begin{flushright}\vskip -15pt 
\rule{0.45\columnwidth}{.1pt} \end{flushright}
\begin{multicols}{2}
Here $4\pi \bar{r}_{s}^{3}/3=\bar{n},$ and $G(r)$ is
normalized to $\tilde{\epsilon}_{xc}(\bar{n})$ instead of $\epsilon _{xc}(%
\bar{n}).$ Since we do {\it not} require that $G(r)=\int_{0}^{1}[g(r;\lambda
,\bar{n})-1]d\lambda .,$ where $g$ corresponds to the uniform gas, but
rather consider it to be a flexible function satisfying two normalization
conditions, further improvement of the method should be possible along the
line described in the previous paragraph, namely the freedom in chosing $G(r)
$ can, and should, be used to yield $K_{xc}$ according to Eq. (\ref{Kxcd})
close to  the linear response of
the homogeneous electron gas, including correct behavior near $q=2k_{F}$. In
Fig. 2, we show $I_{xc}$ calculated according to Eq. \ref{awda} with
\figure{ \centerline{\epsfig{file=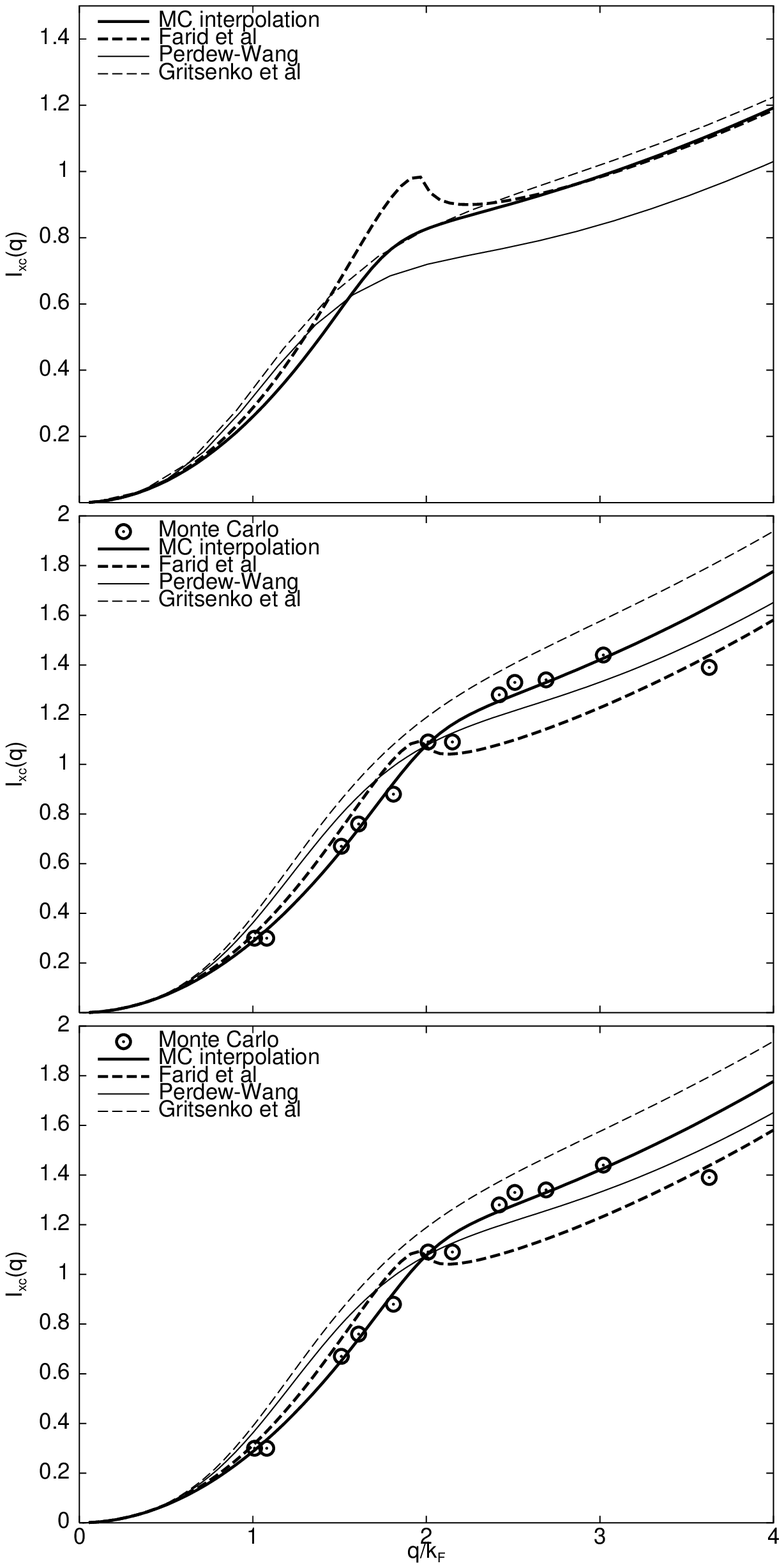,width=0.9\linewidth}}
\vspace{0.1in} \setlength{\columnwidth}{3.2in} \nopagebreak
\caption{$I_{xc}(q)$ as in Fig.1, but for the modified WDA of
Eq.\ref{awda}. Also the analytical formula of Farid et al 
\protect\cite{farid} is shown.}}
the different functional form of $G(r).$ Clearly, the results are much
better than either the  LDA or ``conventional'' WDA. Interestingly, when
the nearly
exact Perdew-Wang function, or exponential function with $k=2,$ are used,
the resulting $I_{xc}(q)$ is close to the analytical function derived by
Farid et al (arguably the best {\it analytically derived} $I_{xc}(q)$
available), while an exponential function with $k=1.5$ is close to the formula
of Ref. \onlinecite{MC}, which is a fit to the Monte Carlo data.

To summarize, we have calculated the exchange-correlation local field
function $K_{xc}$ in the WDA, and found that besides the expected
improvement over the LDA it has two major deficiencies: (1) it does not have
correct asymptotic behavior at $q\rightarrow \infty ,$ and (2) the
characteristic feature at $q\approx 2k_{F}$ is displaced towards smaller $q$%
's. The former can be easily corrected by adding a delta-function
component to $G(r),$ which results in Eq. (\ref{awda}). The
latter is harder to fix, but there are still unused degrees of
freedom in the formalism which may be used to tune the behavior near $%
2k_{F}.$ In our opinion, this new scheme for WDA calculations 
(Eq. \ref{awda}) is currently most promising for practical applications.

This work was supported by ONR.

\end{multicols}

\begin{references}
\bibitem{char1} Ref.\onlinecite{char} is an exception, but note 
that the LDA results
of this work differ from most previous calculation.
\end{references}
\end{document}